\documentstyle[epsfig,aps,twocolumn]{revtex}
\begin{document}
\newcommand{\mb}[1]{\mbox{\boldmath $#1$}}
 
\title{\bf \large
Duality in
the Azbel-Hofstadter problem 
and
the
two-dimensional 
d-wave superconductivity
with a magnetic field
}
\author{Y. Morita$^1$ and
Y. Hatsugai$^{1,2}$}
\address{Department of Applied Physics,
University of Tokyo,
7-3-1 Hongo Bunkyo-ku, Tokyo 113-8656, Japan$^1$ \\
PRESTO, Japan Science and Technology Corporation$^{2}$}
\date{\today}
\maketitle

\begin{abstract}
A single-parameter family of lattice-fermion model is constructed.
It is a deformation of the Azbel-Hofstadter problem
by a parameter $h={\Delta}/t$
(quantum parameter).
A topological number is attached to each energy band.
A duality between 
the classical limit ($h=+0$) and the quantum limit ($h=1$)
is revealed in the energy spectrum and the topological number.

The model has a close relation to 
the two-dimensional d-wave superconductivity
with a magnetic field.
Making use of
the duality and a topological argument,
we shed light on
how the quasiparticles 
with a magnetic field
behave 
especially in the quantum limit.
\end{abstract}
\vskip 1.0cm
\narrowtext


Two-dimensional 
Dirac fermions 
with a gauge field
are of current interest, e.g.,
in the context of the vortex state
in a two-dimensional d-wave superconductivity
\cite{anderson1,gs1,mhs1}.
In our study,
Dirac fermions with a gauge field 
are realized on a two-dimensional lattice.
It is a single-parameter deformation of
the Azbel-Hofstadter problem\cite{azbel1,hofstadter1}.
In this paper,
the parameter $h$
is called 
{\it quantum parameter}\cite{mkm1}.
A topological number is assigned for each energy band
\cite{tknn1,yh1,yy1}.
As the quantum parameter is varied 
continuously
from the classical limit ($h=+0$)
to the quantum limit ($h=1$),
the energy spectrum is reconstructed 
through the change of each topological number 
(plateau transition\cite{yy1,hk1,oshikawa1}).
Although the two limits are {\it not} connected
adiabatically,
we found that
there is a {\it duality}
between the classical and the quantum regime.

The model has a close relation to 
the two-dimensional d-wave superconductivity
with a magnetic field.
Applying the duality and a topological argument,
we provide insights into
the quasiparticle spectrum.
In the quantum limit,
interference effects
become relevant especially at zero energy\cite{mhs1}.
The existence of edge states is discussed as well.
It reflects a non-trivial topology of 
each energy band\cite{yy1,hatsugai1}.

Let us define a key Hamiltonian in our paper,
which is
a single-parameter family of lattice fermion model.
It is a deformation of the Azbel-Hofstadter problem.
The Hamiltonian is 
$
{\cal H}=
\sum_{l,m}
{\bf c}_{l}^{\dagger}
{\cal H}_{lm}
{\bf c}_{m}$
with
\begin{eqnarray}
{\cal H}_{lm}
=
{e}^{iA_{lm}}
\pmatrix{
t_{lm}^{0}&\Delta_{lm}^{0}\cr
{\Delta_{ml}^{0}}^{*}&-t_{ml}^{0}\cr},
\label{gah}
\end{eqnarray}
${\bf c}_{n}^{\dagger}=
\pmatrix{c_{n\uparrow}^{\dagger} c_{n\downarrow}^{\dagger}\cr}$,
${\bf c}_{n}=
\pmatrix{c_{n\uparrow}\cr c_{n\downarrow}\cr}$,
$A_{lm}=-A_{ml}{\in}{\bf R}$ and
$
{\displaystyle \sum_{\put(0,0){\framebox(4,4)}}}
{A_{lm}}/2{\pi}=
{\phi}=
p/q$
($p$ and $q$ are coprime integers and
the summation runs over four links around a plaquette).
Here
$l,m{\in}{\bf Z}^{2}$,
$t_{m{\pm}(1,0),m}^{0}
=t_{m{\pm}(0,1),m}^{0}
=t$,
${\Delta}_{m{\pm}(1,0),m}^{0}
=-{\Delta}_{m{\pm}(0,1),m}^{0}
={\Delta}$
$(t,{\Delta}{\in}{\bf R})$
and the other matrix elements are zero.
$t$ is set to be a unit energy and
a relevant parameter,
{\it quantum parameter},
is defined by $h={\Delta}/t$.
The relation of this model to 
the two-dimensional d-wave superconductivity
with a magnetic field
is discussed later.
In the classical limit $h=+0$,
this model 
decouples to two essentially equivalent Hamiltonians.
It is the Azbel-Hofstadter Hamiltonian
$\sum_{l,m}
{c}_{l}^{\dagger}
{e}^{iA_{lm}}
t_{lm}^{0}
{c}_{m}$.
The energy spectrum at ${\phi}=0$
is given by 
$E={\pm}{\sqrt{A(k)^{2}+|B(k)|^{2}}}$
where 
$A(k)=2t(\cos{k_{x}}+\cos{k_{y}})$,
$B(k)=
2{\Delta}
(\cos{k_{x}}-\cos{k_{y}})$
and $k{\in}[-\pi,\pi]{\times}[-\pi,\pi]$.
The upper and lower bands
touch at four points 
$(\pm{\pi/2},\pm{\pi/2})$
in the Brillouin zone.
The low-lying excitations 
around the gap-closing points
are described by
massless Dirac fermions.

One of the basic observables 
is a topological number for the $n$-th band, ${\cal C}_n$
\cite{tknn1,yh1,yy1}.
It is 
\begin{eqnarray}
{\cal C}_n =\frac 1 {2\pi i} 
\int d{\bf k}\; \hat z \cdot ({\bf{\nabla}}&\times&{\bf A}_n),\;  
 {\bf A}_n = \langle u_n ({\bf{k}}) | {\bf{\nabla}} | u_n({\bf{k}})
 \rangle
\nonumber
\end{eqnarray}
where ${\nabla}={\partial}/{\partial}{\bf k}$ and
$| u_n(\bf{k}) \rangle$ is a Bloch vector 
of the $n$-th band.
The integration $\int d{\bf k}$ 
runs over the Brillouin zone (torus).
The non-zero topological number results in
the existence of edge states.
In order to see it,
put the system on a cylinder and
introduce a fictitious flux through the cylinder
(it is equivalent to 
a twist in the boundary condition)\cite{laughlin1}.
The edge states
move from one boundary to the other
as the fictitious flux quanta $hc/e$ 
is added adiabatically.
The number of carried edge states
coincides with 
the summation of
topological numbers
below the Fermi energy\cite{yy1,hatsugai1}.
Due to the topological stability, 
a singularity 
necessarily occurs
with the change of the topological number 
(plateau transition \cite{yy1,hk1,oshikawa1}).
The singularity is identified with
an energy-gap closing on some points
in the Brillouin zone.




In Figs. 1-3,
the energy spectra are shown.
As $h$ is varied 
continuously
from the classical limit ($h=+0$)
to the quantum limit ($h=1$),
the energy spectrum is reconstructed 
through the plateau transitions.
Although the two limits are not connected
adiabatically,
a main feature in the data
is that
there is a symmetry
between the classical and the quantum regime.
It leads to a claim that

'
{\it
there is a faithful correspondence
between
$\phi=(1/2+p/q)$ in the classical limit $h=+0$
and $\phi=p/q$ in the quantum limit $h=1$}'.

We call this phenomena {\it duality}.
As discussed above,
this model reduces to a doubled Azbel-Hofstader problem
in the classical limit $h=+0$.
The duality is an analogue of a statistical transmutaion
(composite fermion picture)
in the fractional quantum Hall effect\cite{jain1}.
In the composite fermion picture,
a locally attached flux to each fermion
is replaced with a global uniform flux.
In our case,
pairing effects (off-diagonal order)
play the role of shifting a flux globally.
It is reminiscent of a symmetry
between the d-wave pairing and the $\pi$-flux,
in other words, the $SU(2)$ symmetry\cite{azha1}.
Here some comments are in order.
At ${\phi}=0$ and $1/2$, 
we proved analytically that
the claim is exact.
Moreover, as shown in Figs.1 and 3,
topological numbers 
are consistent with the claim.

As an application of the duality,
let us study
the two-dimensional $d$-wave pairing model
with a magnetic field
especially in the quantum limit ($h=1$).
The {\it pairing model} is
$H=
\sum_{l,m}
{\bf c}_{l}^{\dagger}H_{lm}{\bf c}_{m}$
with
\begin{eqnarray}
H_{lm}
=\pmatrix{
t_{lm}&\Delta_{lm}\cr
\Delta_{ml}^{*}&-t_{ml}\cr}.
\label{pairing}
\end{eqnarray}
Under the 
unitary transformation
$c_{n\uparrow}{\rightarrow}d_{n\uparrow}$,
$c_{n\downarrow}{\rightarrow}d_{n\downarrow}^{\dagger}$
(for ${\forall}n$),
it becomes
$H=
\sum_{l,m}
[d_{l\uparrow}^{\dagger}{t_{lm}}d_{m\uparrow}+
d_{l\downarrow}^{\dagger}{t_{lm}}d_{m\downarrow}+
d_{l\uparrow}^{\dagger}{\Delta_{lm}}d_{m\downarrow}^{\dagger}+
d_{m\downarrow}{\Delta_{lm}^{*}}d_{l\uparrow}].$
It is equivalent to
the Bogoliubov-de Gennes Hamiltonian
for the singlet superconductivity.
Here
$t_{lm}^{*}=t_{ml}$ (hermiticity)
and $\Delta_{lm}=\Delta_{ml}$ (SU(2) symmetry)
are imposed as well.
It satisfies a relation
$-(\sigma_{y}H_{lm}\sigma_{y})^{*}
=H_{lm}$, 
which results in a particle-hole symmetry
in the energy spectrum.
The two-dimensional $d$-wave pairing model
with a magnetic field is defined 
by the pairing model (\ref{pairing})
with
\begin{eqnarray}
t_{lm}
&=&{\exp}(iA_{lm})t_{lm}^{0},
\nonumber
\\
{\Delta}_{lm}
&=&{\exp}(i{\varphi}_{lm}){\Delta}_{lm}^{0}
\nonumber
\end{eqnarray}
where
$A_{lm}=-A_{ml},
{\varphi}_{lm}=({\varphi}_{l}+{\varphi}_{m})/2
{\in}{\bf R}$
so that
$t_{lm}^{*}=t_{ml}$ (hermiticity)
and $\Delta_{lm}=\Delta_{ml}$ (SU(2) symmetry)
are satisfied\cite{anderson1,mhs1}.
Performing a gauge transformation \cite{anderson1}
\begin{eqnarray}
{\bf c}_{n}{\rightarrow}
\pmatrix{
{e}^{i{\varphi}_{n}}&0\cr
0&1\cr}
{\bf c}_{n},
\nonumber
\end{eqnarray}
we obtain
\begin{eqnarray}
H_{lm}
=
{e}^{-iA_{lm}}
\pmatrix{
t_{lm}^{0}{e}^{-2iv_{lm}}&{\Delta}_{lm}^{0}{e}^{-iv_{lm}}\cr
{{\Delta}_{ml}^{0}}^{*}{e}^{-iv_{lm}}&-t_{ml}^{0}\cr}
\label{gauged-dp}
\end{eqnarray}
where $v_{lm}=({\varphi}_{l}-{\varphi}_{m})/2-A_{lm}$.
It is a lattice realization of
the Hamiltonian
discussed in ref.\cite{anderson1}
and $v_{lm}$ corresponds to 
the superfluid velocity.
In the case $v_{lm}=0$,
it
reduces to the Hamiltonian (\ref{gah}) and
the quasiparticle spectrum consists of
'Landau levels'
$E=
{\pm}
{\omega}_{H}
\sqrt{n}$ ($n{\in}{\bf N}$)
in the continuum limit
\cite{anderson1}.

In the following,
we set 
a period for $A_{lm}$ and $v_{lm}$
to be $l_{x}{\times}l_{y}$.
Here we emphasize that,
in the context of superconductivity,
$A_{lm}$ and $v_{lm}$ are determined in
a self-consistent way.
Moreover, 
the spatial variation 
of $|{\Delta}_{lm}|$
plays a crucial role especially 
near a vortex core.
In our study, however,
$A_{lm}$ and $v_{lm}$ are treated as adjustable parameters
and
focus is put on the duality or topological arguments
which do not depend on the detail of the potential.


In Fig.4,
an example of the density of states (DoS)
is shown
for the $d$-wave pairing model 
with a magnetic field
in the quantum limit ($h=1$).
Although
the weak-field regime (${\phi}{\sim}0$)
is relevant to reality,
we show the case ${\phi}=1/5$ for clarity.
In the weak-field regime,
the number of energy bands increases but
the following arguments are robust.

As the system at ${\phi}=p/q$ approaches
the quantum regime ($h{\sim}1$),
the quasiparticle spectrum 
becomes close to
that of the Azbel-Hofstadter problem at ${\phi}=(1/2+p/q)$.
The Landau levels are mixed due to
lattice effects and $v_{lm}$.
It causes a singularity at zero energy \cite{mhs1} and
the broadening of each level ('Landau bands').
It is 
due to quantum interference effects
through spatially varying potentials.
It is an analogue of
the vanishing DoS at zero energy
in random Dirac fermions \cite{lfsg1,ntw1,yy2}.
In the case of ${\phi}=p/q$ ($q$=odd),
it is a natural consequence 
of the duality.
In other words,
it can be interpreted by 
the fact 
that there exists a singularity at zero energy 
i.e. $\rho (E=0)=0$
in the Azbel-Hofstadter problem at ${\phi}=(1/2+p/q)$.
It is also to be noted that,
apart from the singularity and the broadening,
the energy spectrum
of the Azbel-Hofstadter problem at ${\phi}=(1/2+p/q)$
$(p=1,q{\gg}1)$
is ${\sim}{\pm}{\omega}_{H}\sqrt{n}$ near the band center
and ${\sim}{\pm}({\omega}{n}+C)$ near the band edge
($n{\in}{\bf N}$).

Now
put the system on a cylinder
(periodic in the $y$ direction and 
open in the $x$ direction)
and let us study 
the basic properties of edge states in
the d-wave pairing model with a magnetic field.
The Schr${\ddot o}$dinger equation 
for the Hamiltonian (\ref{pairing})
is
\begin{eqnarray}
\sum_{j}
\pmatrix{
t_{ij}&{\Delta}_{ij}\cr
{\Delta}_{ji}^{*}&-t_{ji}^{*}\cr}
\pmatrix{u_{j}\cr v_{j}\cr}
=E
\pmatrix{u_{i}\cr v_{i}\cr}.
\nonumber
\end{eqnarray}
Decompose
all the sites ${\cal N}$
into two sublattices $A$ and $B$
($A{\cup}B={\cal N}$
and $A{\cap}B={\phi}$)
where
$t_{ij}$ and ${\Delta}_{ij}$ connecting the same sublattice
are zero.
Define ${\bar w}_{k}$ by
${\bar w}_{k}=+{w}_{k}$ ($k{\in}A$) and 
${\bar w}_{k}=-{w}_{k}$ ($k{\in}B$).
Then it follows from the SU(2) symmetry that
\begin{eqnarray}
\sum_{j}
\pmatrix{
t_{ij}&{\Delta}_{ij}\cr
{\Delta}_{ji}^{*}&-t_{ji}^{*}\cr}
\pmatrix{-{\bar v}_{j}^{*}\cr {\bar u}_{j}^{*}\cr}
=E
\pmatrix{-{\bar v}_{i}^{*}\cr {\bar u}_{i}^{*}\cr}.
\nonumber
\end{eqnarray}
It leads to the claim that,
{\it if an edge state exists},
a paired
(degenerate but linearly independent) 
edge state can be constructed
which is localized in the same side\cite{cmt1}.

Next we shall discuss the existence of edge states.
At first,
set ${\phi}=p/q$, 
$v_{lm}=0$
and focus on
an energy gap with a non-zero topological number 
(see, for example, Fig.3. The duality implies
that a topological number for a generic gap 
is non-zero {\it even} number in the quantum regime.
It is consistent with the previous result \cite{yy1}).
As discussed above,
a non-zero topological number
leads to the existence of edge states
in the gap.
In other words,
the existence of edge states is topologically stable.
Next consider the case when $v_{lm}$ is finite.
As the $v_{lm}$ is varied from zero to the finite value,
the Landau bands are deformed and 
can be overlapped like a semimetal.
We can,
however, 
observe edge states due to the topological stability,
when the plateau transition is absent in the deformation.
In fact, see Fig.5.
The existence of
edge states
can be confirmed
in each energy gap.
It reflects a non-trivial topology 
of each energy band.
In the above discussion,
we have employed a cylinder.
It is possible to consider the same problem
on a geometry with 'defects' (e.g. annulus).
In the case,
the states analogous to the edge states on a cylinder
may be bound to the defects.

In summary,
a single-parameter family of lattice-fermion model is constructed
in two dimensions.
The parameter $h={\Delta}/t$ is called
{\it quantum parameter}.
A {\it duality} 
is revealed in the model
between 
the classical limit ($h=+0$) and the quantum limit ($h=1$).
Employing the duality and a topological argument,
we provide insights into
how the quasiparticles
with a magnetic field 
behave 
especially in the quantum limit.
A more detailed study 
including a self-consistent potential 
and the fluctuation
is left as a future problem.
It is crucial for the analysis of dynamical properties.
As discussed in the context of the integer quantum Hall effect,
when the potential is sufficiently strong,
it can bring the system to a totally different state
\cite{yy3}.

One of the authors (Y.M.)
thanks
H.~Matsumura
for valuable discussions.
This work was supported in part by Grant-in-Aid
from the Ministry of Education, Science and Culture
of Japan.
The computation  has been partly done
using the facilities of the Supercomputer Center,
ISSP, University of Tokyo.


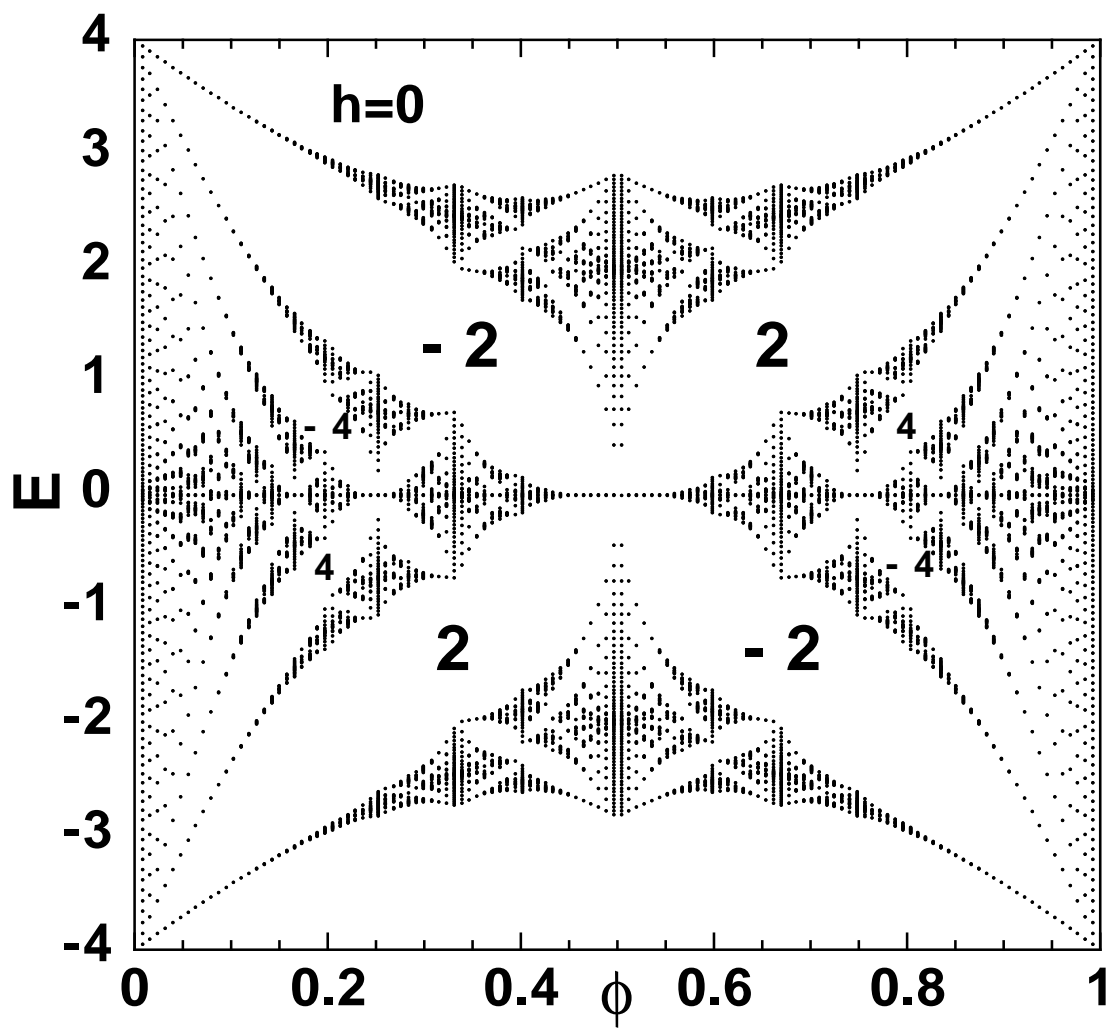
\begin{figure}
\caption{
${\phi}=p/127$, $h={\Delta}/t=0$.
An integer in an energy gap
is the summation of topological numbers 
below the gap.
}
\end{figure}


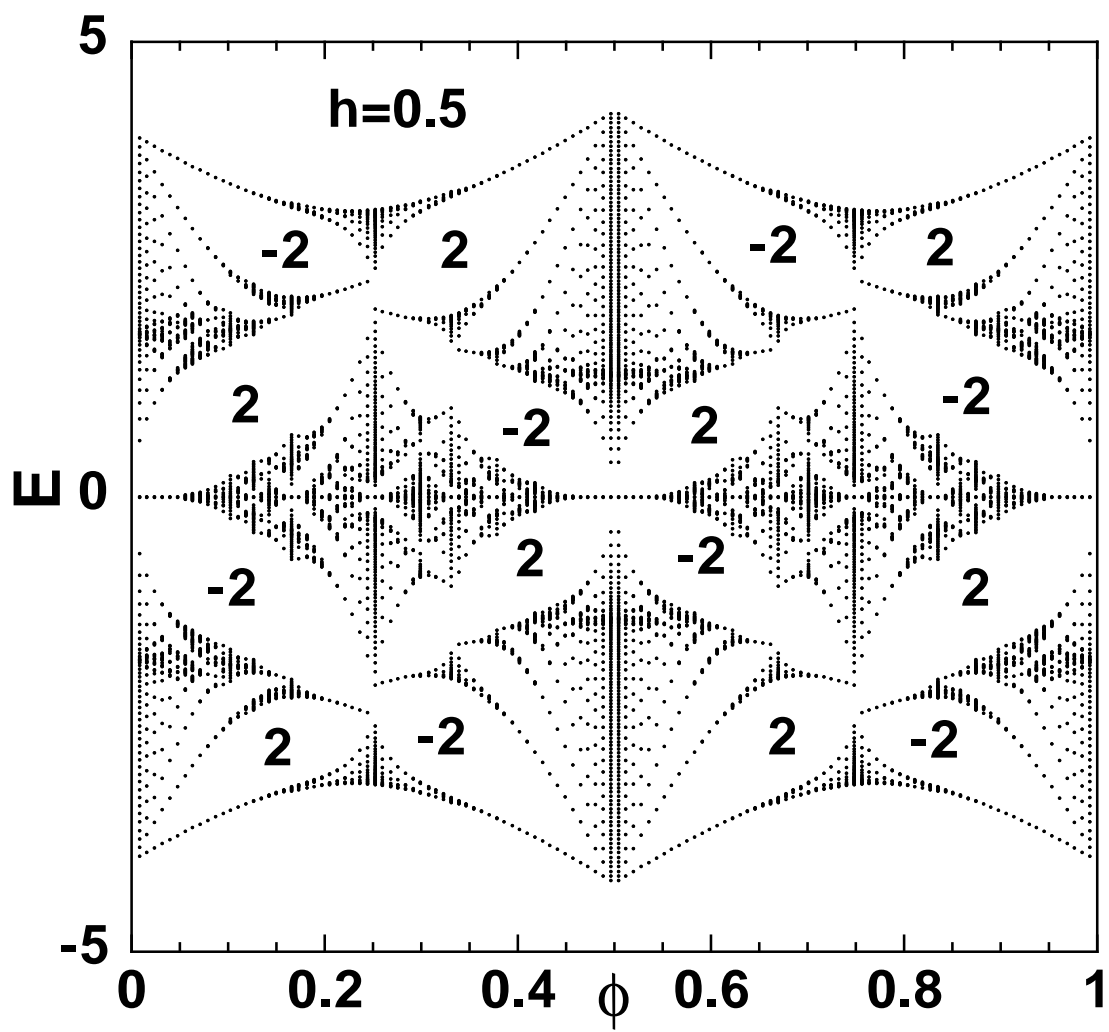
\begin{figure}
\caption{
${\phi}=p/127$, $h={\Delta}/t=0.5$.
An integer in an energy gap
is the summation of topological numbers
below the gap.
}
\end{figure}


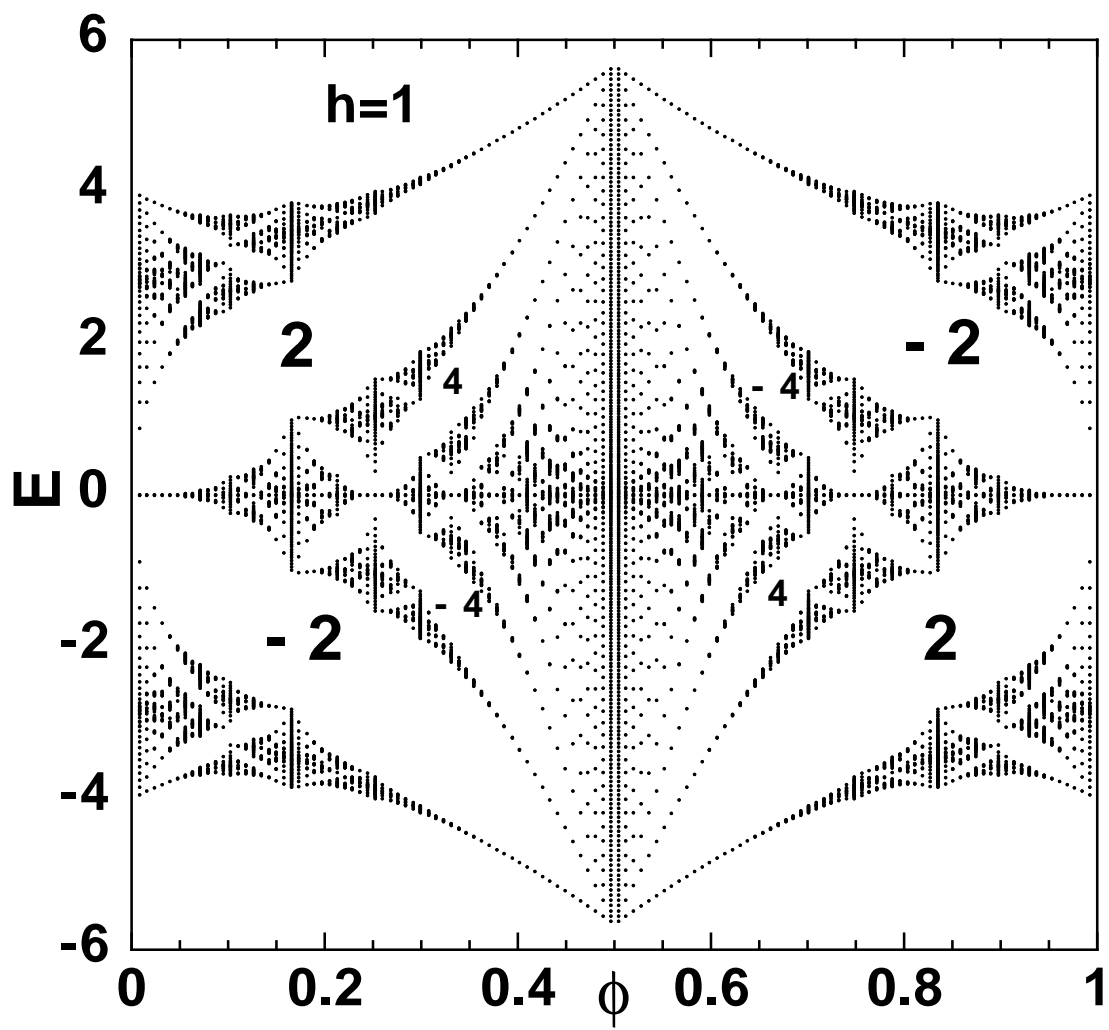
\begin{figure}
\caption{
${\phi}=p/127$, $h={\Delta}/t=1$.
An integer in an energy gap
is the summation of topological numbers 
below the gap.
}
\end{figure}


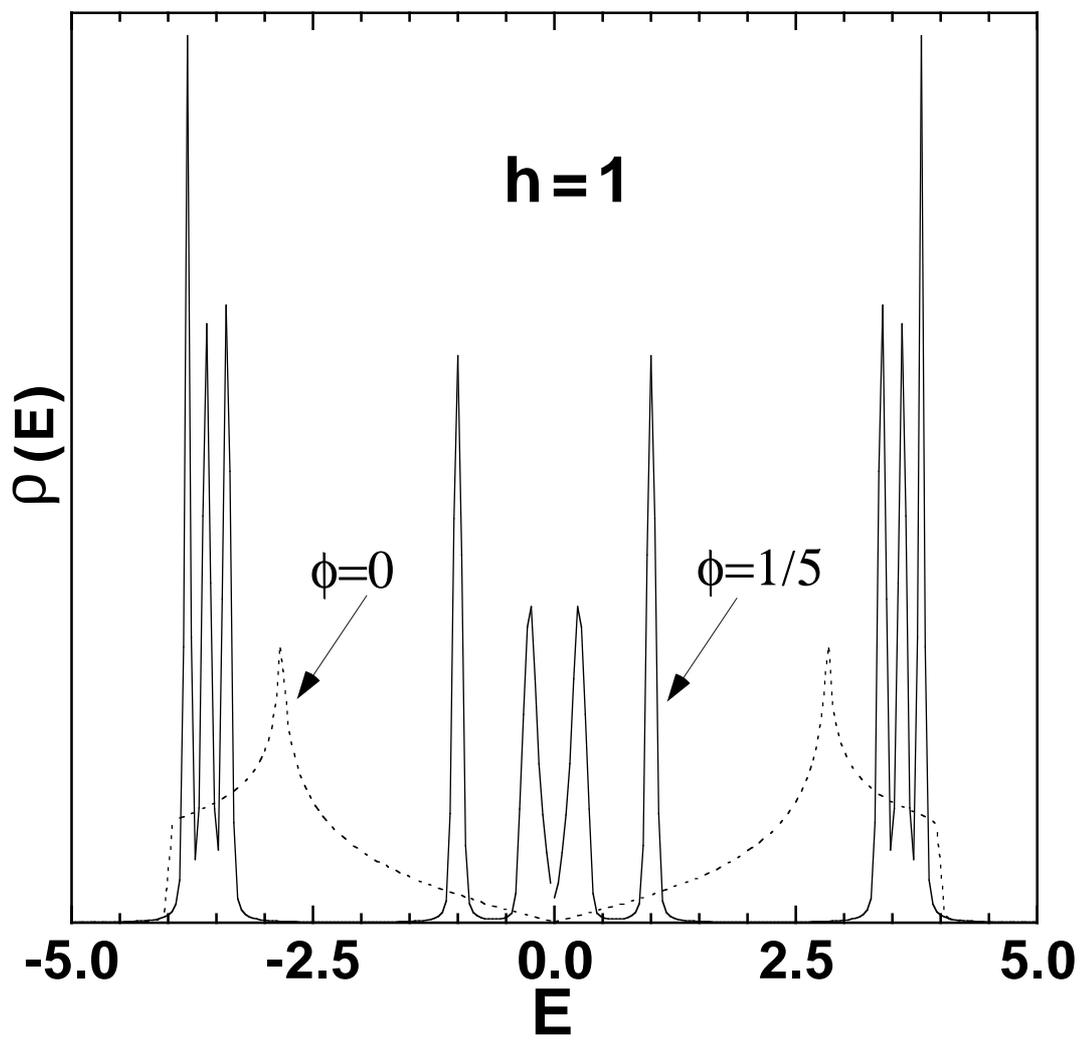
\begin{figure}
\caption{
An example of the DoS
for the $d$-wave pairing model 
($h={\Delta}/t=1$)
with a magnetic field
($p=1$, $q=5$, ${\phi}=1/5$).
The spatial average and variance 
of $v_{lm}$ is $0$ and $0.1$
respectively.
The period
$l_{x}{\times}l_{y}$
is $5{\times}5$.
The DoS
for the $d$-wave pairing model 
($h=1$) without a magnetic field 
is also shown (broken line).
}
\end{figure}


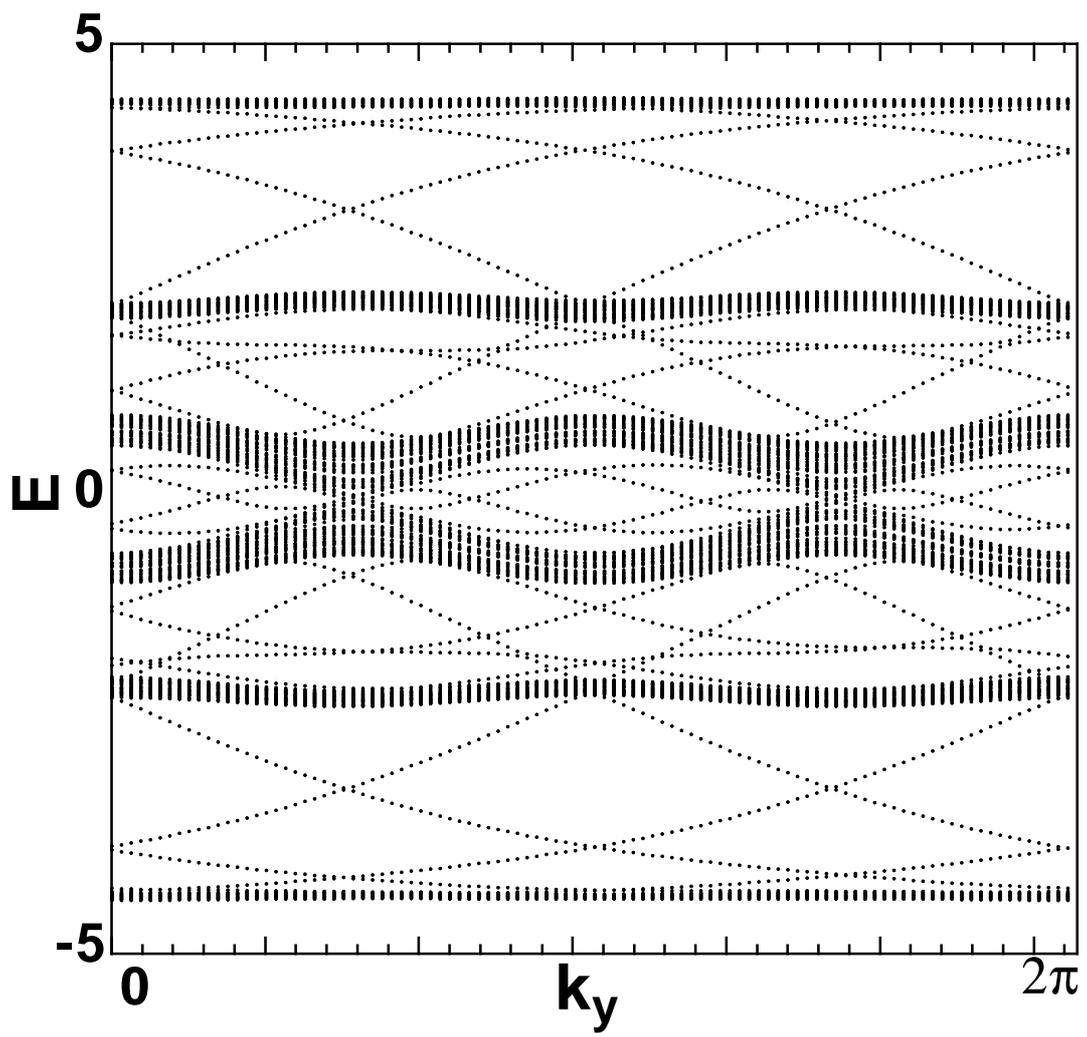
\begin{figure}
\caption{
Energy spectrum
of the $d$-wave pairing model 
($h={\Delta}/t=1$)
with a magnetic field
($p=1$, $q=3$, ${\phi}=1/3$)
on a cylinder 
(the boundary is 
periodic in the $y$ direction and 
open in the $x$ direction).
The spatial average and variance 
of $v_{lm}$ is $0$ and $0.1$ respectively.
The period $l_{x}{\times}l_{y}$
is $30{\times}3$.
The system is composed of $1{\times}100$ cells.
}
\end{figure}



\begin{references}
\bibitem{cmt1}
If the paired edge state is linearly dependent,
$|u_{i}|^{2}+|v_{i}|^{2}=0$ must hold for all sites.
Since
the state is not a null vector,
it leads to a contradiction.
\end{references}
\end{document}